\documentclass[preprint,showpacs,preprintnumbers,superscriptaddress]{revtex4-1}

\usepackage{amsmath,amssymb,graphicx,bm,subfigure,float,siunitx,color}
\usepackage{natbib}
\newcommand{\be}{\begin{equation}}
\newcommand{\ee}{\end{equation}}

\newcommand{\1}{\left}
\newcommand{\2}{\right}
\def\({\left(}
\def\){\right)}
\def\[{\left[}
\def\]{\right]}

\newcommand{\dif}{\,\mathrm{d}}

\newcommand{\m}{\mu}
\newcommand{\n}{\nu}
\newcommand{\al}{\alpha}

\newcommand{\na}{\nabla}

\begin{document}

\title{\boldmath   Stability of the de-Sitter universe: One-loop nonlocal $f(R)$ gravity}

\author{Haiyuan Feng}
\email{Email address: fenghaiyuanphysics@gmail.com }
\affiliation{Department of Physics, Southern University of Science and Technology, Shenzhen 518055, Guangdong, China}

\author{Yi Liao\footnote{Corresponding author}}
\email{Email address: liaoyitianyi@gmail.com}
\affiliation{Department of Physics, College of Mechanical and Electronic Engineering, Fujian Agriculture and Forestry University, Fuzhou, 350002, China}
\affiliation{Department of Materials Science and Engineering, College of Engineering, Southern University of Science and Technology, Shenzhen, 518055, China}

\author{Rong-Jia Yang}
\email{Email address: yangrongjia@tsinghua.org.cn}
\affiliation{College of Physical Science and Technology, Hebei University, Baoding 071002, China}

\begin{abstract}
With the method of the background field expansion, we investigate the one-loop quantization of the Euclidean nonlocal $f(R)$ model in the de-Sitter universe. We obtain the ghost-free condition (GFC) based on the transformation from the Jordan frame to the Einstein frame and the classical stability condition (CSC) satisfied $f^{(0)}_{RR}-\phi_0F^{(0)}_{RR}<0$. 
We present the on-shell and off-shell one-loop effective action and quantum stability condition (QSC) by utilizing the generalized zeta-function. 
We find that under the fulfillment of GFC, CSC and QSC are inconsistent.
\end{abstract}

\maketitle

\section{Introduction}
It is widely acknowledged that strong evidence has recently emerged supporting the accelerated expansion of the universe. This acceleration is attributed to the presence of an effective positive cosmological constant and associated with this dark energy issue. Dark energy offers an explanation for the observed accelerated expansion of the universe. This phenomenon is well-supported by a wide array of astronomical observations \cite{SupernovaCosmologyProject:1998vns, SupernovaSearchTeam:1998fmf, SNLS:2005qlf, SDSS:2003eyi, WMAP:2010qai, SDSS:2004kqt, SDSS:2005xqv, Jain:2003tba}. Observations allow data to accurately constrain cosmic parameters \cite{Kilbinger:2008gk}.  It is remarkable that dark energy aligns with the common assumption that General Relativity (GR) based on  Einstein-Hilbert action serves as the correct theory of gravity.  Nevertheless, one drawback of GR is that it needs to be modified at the ultraviolet (UV) limit, as it fails to adequately explain microscopic physics. The Einstein-Hilbert action leads to non-renormalizability, which has been acknowledged for decades \cite{t1974one,Deser:1974cz,Deser:1974xq}.

The modified gravity models serve as a purely gravitational alternative to explain dark energy. The fundamental concept behind these approaches involves augmenting the gravitational Einstein-Hilbert action with additional gravitational terms. These terms might dominate the cosmological evolution either during the very early or the very late epochs of the universe.
One well explorable possibility in this field was quantum $R^2$ gravity \cite{Buchbinder:1992rb}. It adeptly elucidates the early cosmic inflation phenomenon by introducing additional curvature terms in parallel. Additionally, a straightforward $f(R)$ model that directly extended to the curvature scalar $R$ was investigated, along with its related quantum behaviour \cite{Codello:2007bd,Machado:2007ea,Codello:2008vh,Knorr:2019atm}. This model can provide insights into both the early universe's inflationary phase and its subsequent late-time accelerated expansion. From the perspective of quantum $f(R)$ gravity, initial calculations were performed for the one-loop divergent part in the maximally symmetric spacetime \cite{Cognola:2005de}, and these computations eventually were extended to arbitrary scenarios \cite{Ruf:2017bqx}. Futhermore, another approach to gravity modification was higher-order gravity \cite{Alvarez-Gaume:2015rwa}, which  incorporates contributions from higher-order curvature tensors \cite{Salam:1978fd,Masuda:1976qg,Tomboulis:1983sw}. However, a challenge arises with this model as it gives rise to ghost particles with a mass of spin-$2$, exhibiting the non-unitary in its original quantization following the Feynman prescription \cite{PhysRevD.16.953}. Subsequently, numerous promising works have been tried to address the issue of unitarity \cite{Tomboulis:1977jk, Tomboulis:1980bs, Antoniadis:1986tu}.


One of the most promising theories within modified gravity, which is nonlocal gravity \cite{Modesto:2017sdr, Belgacem:2017cqo, Koshelev:2016xqb}.
The model could guarantee the unitarity and avoid the ghost particles. Simultaneously, the nonlocal interaction term could be adopted in various physical fields, particularly the string theory \cite{Dimitrijevic:2022fhj,Koshelev:2016xqb,Deser:2007jk}. The modification that contains a function of the operator $\Box^{-1}$ has been investigated \cite{Deser:2007jk}. In addition, the history of the universe, including periods such as inflation, radiation/matter dominance, and the dark epoch, could be explained by this model
\cite{Koivisto:2008dh}. It aligns with the results of testing the Solar System and provides a coherent framework for comprehending these cosmic epochs.

In this paper, we will explore the modified nonlocal $f(R)$ model with a particular emphasis on investigating the stability of the de Sitter solution. The stability holds significance in various contexts; for instance, in the $\Lambda$CDM model, it ensures the absence of future singularities in the solution. However, it's worth noting that the cosmological term must suffers the well-known cosmological constant problem, which remains an unsolved issue to date. On the contrary, as previously noted, modified gravity models may offer a naturally geometric approach consistent with Einstein's original concepts. Therefore, the stability or instability around a de-Sitter solution is of some interest in modified gravity.



The paper is organized as follows: In Section II, we will firstly localize nonlocal $f(R)$ gravity by redefining the field, then transform it into the Einstein frame and establish the ghost-free condition (GFC). In Section III, we will employ the mini-superspace approach to derive the perturbation matrix and compute the classical stability condition (CSC) in the de-Sitter spacetime. In Section IV, based on the most fundamental zeta-function method for one-loop calculations, we will investigate the effective action of nonlocal $f(R)$ gravity in the de-Sitter background. We find that the difference between this result and that in GR lies in the contributions from the scalar modes, with its minimum eigenvalue determined by the roots of a quartic equation. Finally, we will explore the quantum stability condition (QSC) in the one-loop context and provide the on-shell and off-shell one-loop divergence term using Vieta's theorem.

\section{Nonlocal $f(R)$ gravity and ghost-free condition }
Nonlocal gravity theories have emerged as a focal point of theoretical research due to their more favorable quantum behavior \cite{Deser:2007jk}. Initially, some nonlocal theories was used to explain the accelerated expansion of the universe and later evolved into a gravitational theory explaining quantum phenomena \cite{NOJIRI2008821, Bamba:2012ky, Zhang:2011uv, Elizalde:2011su, Nojiri:2010pw, Zhang:2016ykx}. The profound influence of the phenomena is revealed through the utilization of a nonlocal interaction term (This nonlocal interaction term is also present in the string/M theory) \cite{Nojiri:2007uq, Jhingan:2008ym}. In an effort to describe physical phenomena, most nonlocal quantum gravity models introduce either nonlocal scalar fields or the d'Alembertian operator $\Box$. In our research, we focus on a nonlocal $f(R)$ model, represented by the following action
\be
\label{1}
S=\int \dif^4x\sqrt{-g}\[f(R)-\frac{1}{2}F(R)\Box^{-1}F(R) \],
\ee
where  $R$ is the Riemann curvature scalar. Both $f(R)$ and $F(R)$ are functions of curvature scalar.  The field equation are nonlinear integro-differential equation due to the
nonlocal term. For convenience, it can be proved that the above action is equivalent to the local form by performing transformation $\phi\equiv\Box^{-1}F(R)$ \cite{Nojiri:2019dio}.
\be
\label{2}
S=\int \dif^4x\sqrt{-g}\[f(R)-\frac{1}{2}\partial_\m\phi\partial^\m\phi-\phi F(R) \],
\ee
such a transformation of nonlocality into locality simplifies the computation of the equation. Additionally, by using two extra auxiliary fields, $A$ and $B$, we can further modify the above action reading
\be
\label{3}
S=\int \dif^4x\sqrt{-g}\[f(A)-\frac{1}{2}\partial_\m\phi\partial^\m\phi-\phi F(A)+B(R-A) \].
\ee
The variation of the aforementioned two fields respectively yield the constraint equations
\be
\1\{\begin{split}
\label{4}
&A=R\\
&B=f_A-\phi F_A,\\
\end{split}\2.
\ee
where $f_A\equiv\frac{\dif f(A)}{\dif A}$  and $F_A\equiv \frac{\dif F(A)}{\dif A}$. In order to provide the ghost-free condition (GFC), we transform it to the Einstein frame by the conformal transformation $g_{\m\n}=e^{-\tau}\tilde{g}_{\m\n}$ and $B\equiv e^{\tau}$,
\be
\1\{\begin{split}
\label{5}
&S=\int \dif^4x\sqrt{-g}\[\tilde{R}-\frac{3}{2}\partial_\mu\tau\partial^{\mu}\tau-\frac{1}{2}e^{-\tau}\partial_\m\phi\partial^\m\phi-U(\phi,\tau) \]\\
&U(\phi,\tau)\equiv-e^{-2\tau}f(A(\phi,\tau))+e^{-\tau}A(\phi,\tau)+\phi e^{-2\tau}F(A(\phi,\tau)).\\
\end{split}\2.
\ee
By examining \eqref{4}, we can ascertain that the field $A$ exhibits a dependence on both $\tau$ and $\phi$.  We can establish the GFC through the analysis of the kinetic term. The condition met when the coefficient determinant is greater than zero. This condition is expressed as $B = f_A(A) - \phi F_A(A) = e^{\tau} > 0$. Consequently, this criterion provides a robust assurance that the nonlocal $f(R)$ gravity theory remains devoid of ghost term.
\section{Classical stability condition}
It is imperative to investigate the classical stability condition (CSC) that the model adheres to under the de-Sitter solution. To achieve this, we will utilize the mini-superspace approach to establish the CSC as outlined in references \cite{Cognola:2007vq, Pozdeeva:2019agu}. The equation of motion for the model can be derived by varying \eqref{2} as follows
\be
\1\{\begin{split}
\label{6}
&\Box\phi=F(R)\\
&g_{\m\n}\[f(R)-\frac{1}{2}\partial_\m\phi\partial^\m\phi-\phi F(R)\]-2R_{\m\n}\(f_R-\phi F_R\)\\
&+2\na_{\m}\na_{\n}\(f_R-\phi F_R\)-2g_{\m\n}\Box\(f_R-\phi F_R\)+\na_{\m}\phi\na_{\nu}\phi=0.  \\
\end{split}\2.
\ee
We find that there exist constant solutions $R = R_0$ and $\phi = \phi_0$, which could construct maximally symmetric de-Sitter solution. By substituting constant solutions into the field equation \eqref{6}, we can obtain
\be
\1\{\begin{split}
\label{7}
&F(R_0)=0\\
&f(R_0)=\frac{1}{2}R_0\(f^{(0)}_R-\phi_0 F^{(0)}_R\), \\
\end{split}\2.
\ee
where $f^{(0)}_R\equiv\frac{\dif f}{\dif R}|_{R=R_0}$ and $F^{(0)}_R\equiv\frac{\dif F}{\dif R}|_{R=R_0}$. It's important to note that $F(R_0)=0$ does not necessarily imply that its derivative is also zero. Therefore, when $R_0$ satisfies condition $F(R_0) = 0$, $F^{(0)}_R\neq0$ and is positive, we have a solution describing de-Sitter spacetime with $\phi_0=\frac{R_0f^{(0)}_R-2f(R_0)}{R_0 F^{(0)}_R}$. Then, the  ghost-free condition $B > 0$ is equivalent to $f(R_0)>0$. when $R_0$ is negative or zero, it is associated with anti-de-Sitter solution and Minkowski spacetime. we primarily focus on the de-Sitter case in the article.


Subsequently,  in order to investigate whether this solution is a stable point of minimum, we redirect our focus to isotropic and homogeneous solution using the spatially flat  FRW metric, specifically
\be
\label{8}
\dif s^2=-N(t)\dif t^2+a^2(t)\(\dif x^2+\dif y^2+\dif z^2\),
\ee
where $t$ represents cosmic time, $N(t)$ is an arbitrary lapse function that exhibits this degree of gauge freedom associated with the reparametrization invariance of the mini-superspace gravitational model. The curvature scalar for the metric tensor can be determined by
\be
\label{9}
R=6\(\frac{\ddot{a}}{aN^2}+\frac{\dot{a}^2}{a^2N^2}-\frac{\dot{a}\dot{N}}{aN^3} \).
\ee
To operate within the first-derivative gravitational system, we introduce the Lagrange multiplier $y$ to express \eqref{2} as
\be
\label{10}
S=\int \dif^3x\int \dif t Na^3\[f(R)-\frac{1}{2}\partial_\m\phi\partial^\m\phi-\phi F(R)-y\(R-6\(\frac{\ddot{a}}{aN^2}+\frac{\dot{a}^2}{a^2N^2}-\frac{\dot{a}\dot{N}}{aN^3} \)\) \],
\ee
when we perform the variation with respect to $R$, we obtain $y = f_R - \phi F_R$. Furthermore, we insert this expression into above equation to derive the Lagrangian, employing the integration by parts,
\be
\label{11}
\begin{split}
L(a,\dot{a},R,\dot{R},\phi,\dot{\phi},N)&=-\frac{6\dot{a}^2a\(f_R-\phi F_R\)}{N}-\frac{6\dot{a}a^2\(\dot{R}\(f_{RR}-\phi F_{RR}\)-\dot{\phi}F_R\)}{N}\\
&+Na^3\(f-Rf_R-\phi\(F-RF_R\)+\frac{\dot{\phi}^2}{2N^2}\).
\end{split}
\ee
In this scenario, the Lagrangian involves $a$, $R$, $\phi$, $N$ and their derivatives as independent variables. Four equations of motion can be derived, with three of them being independent. These three independent equations are used in the analysis system. Specifically, we choice $N(t)=1$ in this research. Subsequently, the equations of motion corresponding to $N$, $R$, and $\phi$ are
\be
\1\{\begin{split}
\label{12}
&\dot{\phi}=Y\\
&\dot{Y}=-F-3HY\\
& \dot{H}=\frac{R}{6}-2H^2\\
&\dot{R}=\frac{Y F_R +\frac{Y^2}{12H}-H\(f_R-\phi F_R\)-\frac{1}{6H}\(f-Rf_R-\phi\(F-RF_R\) \)}{f_{RR}-\phi F_{RR}},
\end{split}\2.
\ee
where $H\equiv\frac{\dot{a}}{a}$ correspond to the Hubble parameter, and the auxiliary field $Y\equiv\dot{\phi}$ can reduce the derivative order of $\phi$. It is worth noting that we don't need to consider the Euler equation for $a$, as it can be derived from the three previously mentioned independent equations.

The critical points $R_0$, $H_0$, which are defined as $\dot{R}=0$ and $\dot{H}=0$, are necessary for the examination of the stability of the system. It is demonstrably true that the right side of the aforementioned formula, which can result in \eqref{7} (the critical point of the de-Sitter solutions). After that, the system is linearized at these points.



\be
\left(
\begin{array}{c}
\delta\dot{\phi}\\
\delta\dot{Y}\\
\delta\dot{R}\\
\delta\dot{H}
\end{array}
\right)
=
\left(
\begin{array}{cccc}
0&1&0&0\\
0&-3H_0&-F^{(0)}_R&0\\
\frac{-H_0F^{(0)}_R}{ f^{(0)}_{RR}-\phi_0F^{(0)}_{RR}}&\frac{F^{(0)}_R}{ f^{(0)}_{RR}-\phi_0F^{(0)}_{RR}}&H_0&\frac{-4f(R_0)}{R_0\(f^{(0)}_{RR}-\phi_0F^{(0)}_{RR}\)}\\
0&0&\frac{1}{6}&-4H_0
\end{array}
\right)
\left(
\begin{array}{c}
\delta\phi\\
\delta Y\\
\delta R\\
\delta H
\end{array}
\right)
\ee
It is straightforward to demonstrate that these two conditions ensure stability. The first requirement automatically guarantees due to the trace of the matrix that is less than zero. The second condition demands that the determinant is greater than zero, which can be equivalently expressed as $f^{(0)}_{RR}-\phi_0F^{(0)}_{RR}<0$. 
 We can also confirmed  the equivalence between CSC investigated through the mini-superspace method and the minima with CSC derived from the \eqref{5}. (potential functions which satisfy $U'=0$ and $U''>0$ in the Einstein framework). However, it's important to note that the matrix degenerates when considering $F(R) = 0$ and $\phi = 0$, thus necessitating additional constraint to obtain the result for the $f(R)$ model.

Given the significant quantum fluctuations during the inflation and accelerating expansion phases, our focus is primarily on the model's quantum behavior. In the following section, we will calculate the one-loop effective action and outline the quantum stability condition (QSC) for the de-Sitter solution.



\section{One-loop effective action and quantum stability condition in the nonlocal $f(R)$ gravity }

In quantum field theory (QFT), the concept of the quantum effective action assumes a pivotal role in comprehending the behavior of quantum fields and their interactions. It emerges as a potent tool for capturing the intricate dynamics inherent in quantum systems. The quantum effective action serves as an extension of the classical action, encompassing quantum fluctuation. It encapsulates the collective influence exerted by all quantum fields, thereby offering profound insights into these influences under various conditions.

One of the effective action fundamental application is calculating the quantum correction, notably through loop diagram. These corrections are crucial in comprehending particle interactions at the quantum level. We can extract valuable information regarding particle masses, coupling constants, and other critical parameters in the correction term. Moreover, the quantum effective action assumes indispensable significance in the examination of phase transitions, notably within the domains of condensed matter physics and cosmology. It facilitates the elucidation of phenomena like spontaneous symmetry breaking and the formation of diverse phases within physical systems \cite{Esposito1997}.

In this section, we employ the background field method to calculate the one-loop effective action of the gravitational field. We will calculate the one-loop partition function in de-Sitter background using the Euclidean (or Wick rotation) approach \cite{Bytsenko:1994bc}. The formulation for the effective action, which applies to the general scalar field case, can be expressed as
\be
\label{8}
e^{-\Gamma^{(1)}}=Z=e^{-S_E[\phi_{cl}]}\int D\varphi e^{-\varphi L  \varphi},
\ee
where $\phi_{cl}$ represents the classical field, $L$ represents fluctuation operator, and $\varphi$ is the perturbation of the background field. $\Gamma^{(1)}$ is the one-loop effective action which contains all the physical information. Taking the logarithm of \eqref{8} gives the expression of $\Gamma^{(1)}$ as
\be
\label{9}
\Gamma^{(1)}=-\ln{Z}=S_E[\phi_{cl}]+\frac{1}{2}\ln\det\frac{L}{\mu^2},
\ee
where $\mu^2$ is a renormalization parameter, which appears in order to make the fluctuation operator dimensionless. The functional determinant can generally be written in integral form
\be
\label{10}
\ln\det\frac{L}{\mu^2}=-\int^{\infty}_{0} \dif t  t^{-1}\text{Tr}e^{-\frac{tL}{\mu^2}},
\ee
one of the most fundamental methods for determining the effective action is the heat kernel trace. We have the asymptotic expansion for a well-defined operator $L$.
\be
\label{11}
\text{Tr}e^{-t\frac{L}{\mu^2}}\approx \sum^{\infty}_{i=0}a_{i}\(\frac{L}{\mu^2}\)t^{i-\frac{d}{2}},
\ee
where $a_{i}\(\frac{L}{\mu^2}\)$ are Seeley-deWitt coefficients \cite{Vassilevich:2003xt}. To address the divergent term in \eqref{10}, we employ zeta-function regularization on the effective action, following the procedure detailed in the literature \cite{Dowker:1975tf}. The expression for the effective action is
\be
\label{12}
\Gamma^{(1)}(\epsilon,\phi_{cl})=S_E[\phi_{cl}]-\frac{1}{2}\int^{\infty}_{0}\dif t \frac{t^{\epsilon-1}}{\Gamma(1+\epsilon)}\text{Tr}e^{-\frac{tL}{\mu^2}}=S[\phi_{cl}]-\frac{1}{2\epsilon}\zeta(\epsilon|\frac{L}{\mu^2}),
\ee
In four {\color{red}dimension}, the function $\zeta(s|L)$ converges for $Re(s) > 2$ and can be expressed as
\be
\1\{\begin{split}
\label{13}
&\zeta(s|L)=\frac{1}{\Gamma(s)}\int^{\infty}_{0} \dif t t^{s-1}\text{Tr}e^{-t L}\\
&\zeta(s|\frac{L}{\mu^2})=\mu^{2s}\zeta(s|L),\\
\end{split}\2.
\ee
where the integral representation of zeta-function is given by the above equation. We can eventually extract the contribution of the divergence term by utilizing \eqref{13} and Taylor expansion of zeta-function
\be
\label{14}
\zeta(\epsilon|L)=\zeta(0|L)+\zeta'(0|L)\epsilon+O(\epsilon^2),
\ee
and
\be
\label{15}
\Gamma^{(1)}(\epsilon,\phi_{cl})=S_E[\phi_{cl}]-\frac{1}{2\epsilon}\zeta(0|L)-\frac{\zeta(0|L)}{2}\ln{\mu^2}-\frac{1}{2}\zeta'(0|L).
\ee

The last three terms of the formula should be summed when multiple fluctuation operators are present. We set $\mu=1$ for the present situation since the appearance of $\mu$ only depends on the renormalization group equation. Importantly, the zeta-function evaluated at zero is responsible for completely addressing the divergent term of the one-loop effective action. In the framework of a one-loop renormalizable theory, the divergence term can be effectively eliminated by introducing bare parameters and scale-dependent physical quantities related to coupling constants. Furthermore, the one-loop effective action can be written as the derivative of the zeta-function at zero, which will be resolved in the following section.

\subsection{Quantum field fluctuations around the de-Sitter background}
We explore the one-loop quantization of the nonlocal $f(R)$ model in the de-Sitter background. Initially, the Euclidean action given by \eqref{2} can be described as
\be
\label{16}
S_E=\int \dif^4x\sqrt{g}\[f(R)-\frac{1}{2}\partial_\m\phi\partial^\m\phi-\phi F(R) \],
\ee
The model has a de-Sitter solution with constant curvature since it satisfies on-shell condition \eqref{7}. The spacetime of positive curvature scalar $R_0$ has the topological structure $S_4$. Its metric tensor is denoted by
\be
\label{17}
\dif S_E^2=\dif t_E^2\(1-H_0 r^2 \)+\frac{\dif r^2}{(1-H_0r^2)}+r^2\dif \Omega^2,
\ee
where $\dif \Omega^2$ represents the line element of the unit sphere. The volume of this metric is
\be
\1\{\begin{split}
\label{18}
&V\(S_4\)=\frac{384\pi^2}{R^2_0}\\
& R_0=12H^2_0,\\
\end{split}\2.
\ee
while Riemann and Ricci tensors are given by the metric and can be characterized as
\be
\1\{\begin{split}
\label{19}
&R^{(0)}_{ijkl}=\frac{R_0}{12}\(g^{(0)}_{ik}g^{(0)}_{jl}-g^{(0)}_{il}g^{(0)}_{jk}\)\\
&R^{(0)}_{ij}=\frac{R_0}{4}g^{(0)}_{ij}.\\
\end{split}\2.
\ee

Subsequently, the metric field can be characterized as the background field adds a perturbed term, and the component can be expressed as
\be
\1\{\begin{split}
\label{20}
&g_{ij}=g^{(0)}_{ij}+h_{ij}\\
&g^{ij}=g^{(0)ij}-h^{ij}+h^{ik}h^{j}_{k}\\
&h=g^{(0)ij}h_{ij},\\
\end{split}\2.
\ee
where, we only consider terms up to second order in the metric perturbation. The following equation is required to expand the action to second-order term
\be
\1\{\begin{split}
\label{21}
&\frac{\sqrt{g}}{\sqrt{g^{(0)}}}=1+\frac{h}{2}+\frac{h^2}{8}-\frac{1}{4}h_{ij}h^{ij}\\
& R^{(1)}=\na^{i}\na^{j}h_{ij}-\Box h-\frac{R^{(0)}}{4}h \\
&R^{(2)}=\frac{1}{4}h_{ij}\Box h^{ij}+\frac{1}{4}h\Box h+\frac{R^{(0)}}{12}h^{ij}h_{ij}+\frac{1}{2}\na_ih^{ij}\na_kh^{k}_{j}+\frac{1}{24}h^2,
\end{split}\2.
\ee
$\nabla_i$ represents the covariant derivative operator associated with the background metric $g^{(0)}_{ij}$. Utilizing equations \eqref{18}-\eqref{21}, we can express the Lagrangian \eqref{16} as follows
\be
\1\{\begin{split}
\label{22}
&S=S_0+S_1+S_2 \\
&S_0=\int\dif^4x \sqrt{g^{(0)}}\mathcal{L}_{0}=\int\dif^4x \sqrt{g^{(0)}}\(f(R_0)-\phi_0F(R_0)\)\\
&S_1= \int\dif^4x \sqrt{g^{(0)}}\mathcal{L}_{1}=\int\dif^4x \sqrt{g^{(0)}}\[\frac{h}{2}\(f(R_0)-\phi_0F(R_0)-\frac{R_0}{2}\(f^{(0)}_{R}-\phi_0F^{(0)}_{R} \) \)-\varphi F(R_0) \] \\
&S_2= \int\dif^4x \sqrt{g^{(0)}}\mathcal{L}_{2},
\end{split}\2.
\ee
The information obtained from $\mathcal{L}_{1}$ pertains to the on-shell condition of the field equation, while $\mathcal{L}_{2}$ provides the analytical expression for the inverse propagator associated with the Lagrangian density. Once all boundary terms have been removed, the final expression for $\mathcal{L}_2$ can be presented as
\be
\1\{\begin{split}
\label{23}
&\frac{1}{2}\(f^{(0)}_{RR}-\phi_0F^{(0)}_{RR}\)h^{ij}\na_{i}\na_{j}\na_{k}\na_{l}h^{kl}-\frac{1}{2}\(f^{(0)}_{R}-\phi_0F^{(0)}_{R}\)h^{ij}\na_i\na_kh^{k}_{j}\\
&+h^{ij}\[ \frac{1}{4}\(f^{(0)}_{R}-\phi_0F^{(0)}_{R} \)\Box-  \frac{1}{4}\(f(R_0)-\phi_0F(R_0) \) +\frac{R_0}{12}\( f^{(0)}_{R}-\phi_0F^{(0)}_{R}\) \]h_{ij}\\
&-h\[ \(f^{(0)}_{RR}-\phi_0F^{(0)}_{RR}\)\Box + \frac{R_0}{4}\( f^{(0)}_{RR}-\phi_0F^{(0)}_{RR}\)-\frac{1}{2}\( f^{(0)}_{R}-\phi_0F^{(0)}_{R}  \)    \]\na^{i}\na^{j}h_{ij}\\
&+h\[ \frac{1}{2}\(f^{(0)}_{RR}-\phi_0F^{(0)}_{RR} \)\Box^2+ \(\frac{R_0}{4}\(f^{(0)}_{RR}-\phi_0F^{(0)}_{RR}\)  -\frac{1}{4}\(f^{(0)}_{R}-\phi_0F^{(0)}_{R}\)   \) \Box\right.\\
&\left.+\frac{1}{8}\(  f(R_0)-\phi_0 F(R_0) \)-\frac{R_0}{12}\(f^{(0)}_{R}-\phi_0F^{(0)}_{R}\)+\frac{R_0^2}{32}\(f^{(0)}_{RR}-\phi_0F^{(0)}_{RR} \)  \] h.
\end{split}\2.
\ee
In the context of the de-Sitter background, it is imperative to maintain second-order perturbations for the influence of diverse modes on one-loop quantum effects. Consequently, the decomposition of the perturbation field $h_{ij}$ becomes a requisite step. In accordance with established practice, we perform a decomposition of $h_{ij}$ into its irreducible components, as described in \cite{Fradkin:1983mq}. This decomposition is typically carried out by decomposing $h_{ij}$ as follows
\be
\label{24}
h_{ij}=\hat{h}_{ij}+\na_i\xi_j+\na_j\xi_i+\na_i\na_j\sigma+\frac{1}{4}g_{ij}\(h-\Box\sigma\),
\ee
where $h$ and $\sigma$ denote the trace and scalar components of tensor, respectively. The vector and tensor modes $\xi$ and $\hat{h}_{ij}$ need to satisfy the transverse traceless condition
\be
\1\{\begin{split}
\label{25}
&\na_i\xi^{i}=0 \\
&\na_i\hat{h}^{ij}=0\\
&\hat{h}^{i}_{i}=0. \\
\end{split}\2.
\ee
According to the above decomposition, we substitute \eqref{24} into \eqref{23} and finally simplify $\mathcal{L}_{2}$ to
\be
\label{26}
\mathcal{L}_{2}=\hat{h}_{ij}O_1\hat{h}_{ij}+\xi_i O_2\xi_i+\sigma O_3\sigma+h O_4h+\varphi O_5\varphi+2\varphi O_6h+2\varphi O_7\sigma+2hO_8\sigma,
\ee
with
\be
\1\{\begin{split}
\label{27}
&O_1=\frac{1}{4}\(f^{(0)}_{R}-\phi_0F^{(0)}_{R}\) \Box_{(2)}-\frac{1}{4}\(f(R_0)-\phi_0 F(R_0)\) +\frac{R_0}{12}\( f^{(0)}_{R}-\phi_0F^{(0)}_{R}\)\\
&O_2=\frac{1}{4}\[2\( f(R_0)-\phi_0 F(R_0) \)-R_0\( f^{(0)}_{R}-\phi_0F^{(0)}_{R} \) \]\(\Box_{(1)}+\frac{R_0}{4}\)\\
&O_3=\frac{3}{32}\Box_{(0)}\(\Box_{(0)}+\frac{R_0}{3}\)\[ 3\(f^{(0)}_{RR}-\phi_0F^{(0)}_{RR}\)\Box_{(0)}\(\Box_{(0)}+\frac{R_0}{3}\)-2\( f(R_0)-\phi_0F(R_0) \)\right.\\
&\left.-\( f^{(0)}_{R}-\phi_0F^{(0)}_{R}\)\(\Box_{(0)}-R_0\)  \]\\
&O_4=\frac{1}{32}\[9\(f^{(0)}_{RR}-\phi_0F^{(0)}_{RR}\)\(\Box_{(0)}+\frac{R_0}{3}\)^2-3\(f^{(0)}_{R}-\phi_0F^{(0)}_{R}\) \(\Box_{(0)}+\frac{R_0}{3}\)\right.\\
&\left.-R_0\(f^{(0)}_{R}-\phi_0F^{(0)}_{R}\)+2\( f(R_0)-\phi_0 F(R_0)\)  \]\\
&O_5=\frac{1}{2}\Box_{(0)}\\
&O_6=-\frac{1}{2}\[\frac{F(R_0)}{2}-\frac{R_0}{4}F^{(0)}_{R}-\frac{3F^{(0)}_R}{4}\Box_{(0)}   \]\\
&O_7=-\frac{3F^{(0)}_{R}}{8}\Box_{(0)}\[\Box_{(0)}+\frac{R_0}{3} \]\\
&O_8=\frac{3}{32}\Box_{(0)}\(\Box_{(0)}+\frac{R_0}{3}\)\[ \(   f^{(0)}_{R}-\phi_0F^{(0)}_{R}  \)-3\( f^{(0)}_{RR}-\phi_0F^{(0)}_{RR} \)\(  \Box_{(0)}+\frac{R_0}{3}  \)  \],\\
\end{split}\2.
\ee
where $\Box_{(0)}$, $\Box_{(1)}$, and $\Box_{(2)}$ represent the Laplace-Beltrami operators acting on scalars, traceless-transverse vector fields, and tensor fields, respectively. In the case of a general gravitational field, the diffeomorphism transformation is analogous to the gauge transformation. A general gauge theory necessitates both a gauge-fixed term and a compensation term, often referred to as the ghost term. Therefore, 
we introduce a gauge condition parameterized by a real parameter denoted as $\rho$.
\be
\label{28}
\chi_{k}=\na_ih^{ik}-\frac{1+\rho}{4}\na_k h,
\ee
and gauge fixed term
\be
\label{29}
\mathcal{L}_{gf}=\frac{1}{2}\chi_{i}G^{ij}\chi_{j}, G_{ij}=\alpha g_{ij},
\ee
the harmonic gauge corresponding to the choice $\rho=1$. The ghost term depends on $\chi_{k}$ can be shown as
\be
\1\{\begin{split}
\label{30}
&\mathcal{L}_{gh}=B^{i}G_{ij}\frac{\delta\chi^{j}}{\delta\epsilon^{k}}C^{k}\\
& \frac{\delta\chi^{j}}{\delta\epsilon^{k}}=g_{ij}\Box+R_{ij}+\frac{1-\rho}{2}\na_{i}\na_{j},\\
\end{split}\2.
\ee
where $C_{k}$ and $B_{k}$ are ghost and anti-ghost vector fields respectively. $\delta\chi^{k}$ is determined by infinitesimal gauge transformation. Neglecting total derivatives, one has
\be
\label{31}
\mathcal{L}_{gh}=\alpha B^{k} \(\Box_{(1)}+\frac{R_0}{4}\)C_{k},
\ee
Similarly, we perform irreducible decomposition of ghost and anti-ghost vector fields respectively
\be
\1\{\begin{split}
\label{32}
&B_{k}=\hat{B}_{k}+\na_{k}b,  \na_{i}\hat{B}^{i}=0, \\
&C_{k}=\hat{C}_{k}+\na_{k}c,  \na_{i}\hat{C}^{i}=0 ,\\
\end{split}\2.
\ee
under this decomposition, the \eqref{30} and \eqref{32} can be derived as
\be
\1\{\begin{split}
\label{33}
& \mathcal{L}_{gh}=\alpha\[\hat{B}^{i}\(\Box_{(1)}+\frac{R_0}{4}\)\hat{C}_{i}+\frac{\rho-3}{2}b\Box_{(0)}\(\Box_{0}-\frac{R_0}{\rho-3} \)c \]\\
&\mathcal{L}_{gf}=\frac{\alpha}{2}\[\xi^{i} \(\Box_{(1)}+\frac{R_0}{4}\)^2\xi_{i}+\frac{3\rho}{8}h\Box_{(0)}\(\Box_{(0)}+\frac{R_0}{3} \)\sigma \right.\\
&\left.-\frac{\rho^2}{16}h\Box_{(0)}h-\frac{9}{16}\sigma\Box_{(0)}\( \Box_{(0)}+\frac{R_0}{3}\)^2 \sigma  \] .
\end{split}\2.
\ee
The total Lagrangian density $\mathcal{L}_{tot}=\mathcal{L}_{2}+\mathcal{L}_{gh}+\mathcal{L}_{gf}$ and appears in the scheme of quantum gravity.  We will calculate the one-loop contribution in the following section through functional integration.
\subsection{One-loop quantum corrected nonlocal $f(R)$ gravity}
In this section, we aim to expand our investigation by calculating the one-loop effective action. We consider the determinant within the path integral, which arises from the alteration of variables during field irreducible decomposition. This procedure can be expressed as follows \cite{Buchbinder:1992rb,Fradkin:1983mq}
\be
\label{34}
\begin{split}
Z^{(1)}&=\det{G_{ij}}^{-\frac{1}{2}}\int D[h_{ij}] D[C_{k}] D[B^k] e^{-\int\dif^4x\sqrt{g}\mathcal{L}_{tot}}\\
&=\det{G_{ij}}^{-\frac{1}{2}}\det{J_1}^{-1}\det{J_2}^{\frac{1}{2}}\int D[h]D[\hat{h}_{ij}]D[\hat{\xi}^{j}]D[\sigma]D[\hat{C}_k]D[\hat{B}^k]D[c]D[b] e^{-\int\dif ^4x\sqrt{g}\mathcal{L}_{tot}},
\end{split}
\ee
The origin of  $\det{G_{ij}}$ from vector field constraint is trivial and doesn't contribute any dynamical degrees of freedom. $J_{1}$ and $J_2$ are caused by variable changes made by ghost and tensor part \cite{Shahidi:2018smw,Fradkin:1983mq}, which can be derived as
\be
\label{35}
J_1=\Box_{(0)}, J_2=\( \Box_{(1)}+\frac{R_0}{4} \)\( \Box_{(0)}+\frac{R_0}{3}\)\Box_{(0)}.
\ee
In the process of computing the one-loop effective action, we need to take into account all possible quantum fluctuations of the fields in the theory.
In order to correctly compute the contribution of quantum fluctuations, we need to eliminate these unphysical modes from the calculation. This is usually done by fixing a gauge, which amounts to choosing a specific way of describing the theory that eliminates the unphysical modes. Various gauge choices can result in distinct expressions for the effective action, yet they are expected to yield the same physical predictions. To obtain the partition function, we now substitute all contributions into \eqref{34} in the Landau gauge, $\rho=1,\alpha\rightarrow\infty$.
\be
\label{36}
\begin{split}
e^{-\Gamma^{(1)}_{off-shell}}&=e^{-S_E[\phi_{cl}]}Z^{(1)}=e^{-S_E[\phi_{cl}]}\det\(-\Box_{(1)}-\frac{R_0}{4}\)^{\frac{1}{2}}  \det\(-\Box_{(0)}-\frac{R_0}{2}\)\\
&\det\(-\Box_{(2)}-\frac{R_0}{3}+\frac{f(R_0)-\phi_0F(R_0) }{ f^{(0)}_{R}-\phi_0F^{(0)}_{R}} \)^{-\frac{1}{2}}\\
&\det\(\(-\Box_{(0)}+s_1\) \(-\Box_{(0)}+s_2\)\(-\Box_{(0)}+s_3\)\(-\Box_{(0)}+s_4\)\)^{-\frac{1}{2}},\\
\end{split}
\ee
and $s_i$ $(i=1,2,3,4)$ is the root of the quartic equation
\be
\1\{\begin{split}
\label{37}
&c_1s^4+c_2s^3+c_3s^2+c_4s+c_5=0\\
&c_1=-12\( f^{(0)}_{RR}-\phi_0F^{(0)}_{RR} \)\\
&c_2=4\(f^{(0)}_{R}-\phi_0F^{(0)}_{R} \)-16R_0\(f^{(0)}_{RR}-\phi_0F^{(0)}_{RR} \)+12(F^{(0)}_R)^2\\
&c_3=-4\( f(R_0)-\phi_0F(R_0)\)+6R_0\(f^{(0)}_{R}-\phi_0F^{(0)}_{R}\)  -7R^2_0\( f^{(0)}_{RR}-\phi_0F^{(0)}_{RR}\)\\
&+4F^{(0)}_{R}\(4R_0-6F(R_0) \) \\
&c_4=-2R_0\(f(R_0)+\phi_0F(R_0)\)+2R^2_0\( f^{(0)}_{R}-\phi_0F^{(0)}_{R} \)+12(F(R_0))^2\\
&-20R_0F(R_0)F^{(0)}_R+7R^2_0(F^{(0)}_{R})^2-R^{3}_0 \( f^{(0)}_{RR}-\phi_0F^{(0)}_{RR} \) \\
&c_5=4R_0(F(R_0))^2-4R^2_0F(R_0)F^{(0)}_R+R^{3}_0(F^{(0)}_{R})^2    .
\end{split}\2.
\ee
The result of the one-loop calculation, as indicated by the formula above, is exceptionally intricate. Fortunately, when we incorporate the on-shell condition \eqref{7}, the result can be simplified as
\be
\label{38}
\begin{split}
\Gamma^{(1)}_{on-shell}&=\frac{384\pi^2}{R^2_0}f(R_0)+\frac{1}{2}\ln\det\(-\Box_{(2)}+\frac{R_0}{6} \)-\frac{1}{2}\ln\det\(-\Box_{(1)}-\frac{R_0}{4} \)\\
&+\frac{1}{2}\ln\det\(-\Box_{(0)}+s_+ \)+\frac{1}{2}\ln\det\(-\Box_{(0)}+s_- \)\\
&=\frac{384\pi^2}{R^2_0}f(R_0)  -\frac{1}{2}\zeta'_{\al_2}(0|L_2)+\frac{1}{2}\zeta'_{\al_1}(0|L_1)-\frac{1}{2}\zeta'_{\al_+}(0|L_+)-\frac{1}{2}\zeta'_{\al_-}(0|L_-),
\end{split}
\ee
where
\be
\1\{\begin{split}
\label{39}
&L_2\equiv\(-\Box_{(2)}+\frac{R_0}{6}\), \al_2=\frac{17}{4}+q_2=\frac{9}{4},q_2=-2\\
&L_1\equiv\(-\Box_{(1)}-\frac{R_0}{4}\),   \al_1=\frac{13}{4}+q_1=\frac{25}{4},q_1=3\\
&L_{\pm}\equiv\(-\Box_{(0)}+ s_{\pm}\),\al_{\pm}=\frac{9}{4}+ q_{\pm},q_{\pm}=-\frac{12}{R_0}s_{\pm},
\end{split}\2.
\ee
with
\be
\1\{\begin{split}
\label{40}
&s_{\pm}=\frac{-\lambda_1}{2\lambda_3}\pm\frac{1}{2}\sqrt{\(\frac{\lambda_1}{\lambda_3}\)^2-\frac{4\lambda_2}{\lambda_3}}\\
&\lambda_1\equiv \(f^{(0)}_R-\phi_0 F^{(0)}_R\)+3 \(F^0_{R}\)^2-R_0\(f^0_{RR}-\phi_0F^0_{RR}\)\\
&\lambda_2\equiv R_0(F^{(0)}_{R})^2\\
&\lambda_3\equiv-3\( f^{(0)}_{RR}-\phi_0F^{(0)}_{RR} \).
\end{split}\2.
\ee

For details about the generalized zeta-function's definition and the accurate solution technique, see Appendix A. Within the context of the one-loop effect, the tensor and vector modes are results of general relativity. It is noteworthy that the eigenvalues of Laplacian operators acting on $S^4$ have been extensively documented, as evidenced in \cite{Cognola:2005de}. Simultaneously, in subsection D, we further calculate the on-shell and off-shell one-loop divergent term of nonlocal $f(R)$ gravity in the de-Sitter spacetime. These divergence terms will enhance our comprehension of the model's quantum behavior.

\subsection{Quantum stability analysis}
It is interesting to investigate the region where curvature is small. Furthermore, when curvature is minimal, one may disregard higher powers of curvature in the one-loop effective action, assuming that logarithmic terms play a dominant role. We have the Lagrangian's one-loop correction with the Coleman-Weinberg quantum correction, namely
\be
\label{41}
\begin{split}
L^{(1)}_{eff}\approx R^2\(c_1+c_2\ln{\frac{R}{12}}\)
\end{split}
\ee
where $c_1$ and $c_2$ are constant. We can clearly see from the modified effective Lagrangian density that the GFC and CSC remain unchanged under quantum correction since the quantum corrected action resides within the framework of the classical action. However, we aim to analyze whether the QSC is consistent with the CSC. Equation \eqref{38} serves as a tool for examining QSC with respect to arbitrary perturbations. In this scenario, it is necessary for the operator's eigenvalue to remain non-negative, thereby imposing constraints on the model parameters. In the Appendix A, we derive the eigenvalues of different fluctuation operators. Specifically, the minimum eigenvalues of the Laplacian operators $-\Box_{(0)}$, $-\Box_{(1)}$, and $-\Box_{(2)}$, which correspond to scalar, vector, and tensor fields, are 0, $\frac{R}{4}$, and $\frac{2R}{3}$, respectively. Since the minimum eigenvalues of both tensor and vector operators are non-negative, the focus of the QSC shifts towards the examination of scalar modes. This is equivalent to demanding non-negativity of the roots of the scalar operator, which yield
\be
\1\{\begin{split}
\label{42}
&\frac{\lambda_1}{\lambda_3}<0\\
&\(\frac{\lambda_1}{\lambda_3}\)^2>\frac{4\lambda_2}{\lambda_3}>0.
\end{split}\2.
\ee
It can be demonstrated that this condition is not equivalent to $f^{(0)}_{RR}-\phi_0F^{(0)}_{RR}<0$, which inconsistent with the conclusion presented in the third section. Specifically, when $\phi_0 = F(R_0) = 0$, this result reverts to the stability condition of the $f(R)$ model.

\subsection{The divergence term and renormalization analysis  }
The divergence term manifests as quantum fluctuation inducing infinite contributions, then posing challenge to the theoretical framework. To address this issue, the renormalization group emerges as a potent tool. It facilitates a systematic analysis and control of divergence by introducing appropriate counterterm, ensuring the convergence of physical prediction. In this subsection, we will undertake a comprehensive examination of one-loop divergence and the renormalizability of the theory. To determine the connection between coupling constants and energy scale, we shall consider one-loop divergence term using the zeta-function. From \eqref{15}, it can be concluded that

\be
\label{43}
\begin{split}
\Gamma^{(div)}_{off-shell}(\mu,\epsilon)&=\frac{1}{2\epsilon}\[\zeta_{\al_1}(0,L_1)+2\zeta_{\al_0}(0,L_0)-\zeta_{\bar{\al}_2}\( 0,\bar{L}_2\) -\sum^{4}_{i=1}\zeta_{\al_{s_i}}\( 0,L_{s_i}\)   \].\\
 \end{split}
\ee

Using Vita theorem about the quartic polynomial equation, we reduce the divergence term, which follows
\be
\label{44}
\begin{split}
\Gamma^{div}_{off-shell}&=\frac{1}{2\epsilon}\[-\frac{433}{45}-N_{tot}-\frac{20 (f(R_0)- \phi_0F(R_0) ) \left(3 f(R_0)-3\phi_0 F(R_0) -R_0( f^{(0)}_R- \phi_0  F^{(0)}_R \right)}{R^2_0 \left(f^{(0)}_R-\phi_0  F^{(0)}_R\right){}^2}\right.\\
&\left.-\frac{4 \left(f^{(0)}_R+F^{(0)}_R \left(3 F^{(0)}_R-\phi_0 \right)\right){}^2}{3 R^2_0 \left(f^{(0)}_{RR}-\phi_0 F^{(0)}_{RR}\right){}^2}+\frac{8 f(R_0)-8 F(R_0) \left(\phi_0 -6 F^{(0)}_R\right)+4 R_0 (F^{(0)}_R)^2}{R^2_0 \left(f^{(0)}_{RR}-\phi_0  F^{(0)}_{RR}\right)}\],
\end{split}
\ee
$N_{tot}$ is the total number of zero-modes. When the on-shell condition \eqref{7} under the de-Sitter solution be considered, the divergence term reduces to
\be
\label{45}
\Gamma^{div}_{on-shell}=\frac{1}{2\epsilon}\[-\frac{658}{45}-N_{tot}+\frac{4\(f^{(0)}_{R}+F^{(0)}_{R}\(F^{(0)}_{R}-\phi_0\)\)}{R_0\(f^{(0)}_{RR}-\phi_0F^{(0)}_{RR}\)}-\frac{4\(f^{(0)}_{R}+F^{(0)}_{R}\(3F^{(0)}_{R}-\phi_0\)\)^2}{3R^2_0\(f^{(0)}_{RR}-\phi_0F^{(0)}_{RR}\)^2  } \].
\ee

Not all examples can achieve one-loop renormalizable, and a corresponding one-loop counterterm is required to ensure renormalization. To satisfy the requirement of the de-Sitter solution, we consider the general form of the solution as $F(R)=Q(R)(R-R_0)$ $(Q(R_0)\neq0)$. Substituting this into the on-shell condition \eqref{7}, we obtain $f(R_0)=\frac{1}{2}R_0\(f^{(0)}_R-\phi_0 Q(R_0)\)$. Unfortunately, we can demonstrate that irrespective of the specific forms of $f(R)$ and $Q(R)$, the divergence term separated from the classical action could not cancel the divergence term at the one-loop level. In other words, this model is non-renormalizable in the one-loop calculation.

\section{conclusion and discussion}
In this work, based on the background field, the one-loop quantum correction about the nonlocal $f(R)$ gravity model has been investigated. The de-Sitter solution of the model was taken into consideration as a classical background. We firstly described the Einstein framework for nonlocal $f(R)$ gravity and discovered that the model is ghost-free only in the case of the auxiliary field $B>0$. Further, we used the mini-superspace approach to analyse the classical stability constraint. Finally, we performed the one-loop calculation  in the Euclidean sector, with the background being $S_4$.  The Landau gauge was used by us during the calculation. Subsequently, we derived the quantum stability condition $\frac{\lambda_1}{\lambda_3}<0,\(\frac{\lambda_1}{\lambda_3}\)^2>\frac{4\lambda_2}{\lambda_3}>0$. 
The derived conclusion is that classical stability condition(CSC) and quantum stability condition(QSC) are inconsistent under the fulfillment of ghost-free condition(GFC). Moreover, employing Veda's theorem, we calculate the one-loop divergence term.

In future work, our results can be expanded upon and applied in various contexts. We will use this scheme to analyze the one-loop action in the cosmology. Subsequently, we will employ the quantum corrected conclusion to explore the correspondence with the holographic principle. Furthermore, we will also consider one-loop calculation in a black hole background to investigate alterations in the geometric properties of black holes.

\begin{acknowledgments}
H. Feng would thank Prof. Leonardo Modesto for helpful discussions. Y. Liao would thank Prof. Miao Li. This work is supported in part by National Natural Science Foundation of China under grants (12175099).
This study is also supported in part by National Natural Science Foundation of China (Grant No. 12333008) and Hebei Provincial Natural Science Foundation of China (Grant No. A2021201034).
\end{acknowledgments}

\appendix

\section{The calculation of functional determinants}
In this appendix, we will outline the specific steps for performing one-loop calculation. Following the approach described in \cite{Cognola:2005de}, we initially consider the compact $D$-dimensional manifold as the operational space for the second-order differential operator. Consequently, we give the generalized zeta-function ($\text{Re } s > \frac{D}{2}$) as follows
\be
\1\{\begin{split}
\label{A1}
&\hat{\zeta}(s|L)\equiv\sum_{n}\hat{\lambda}^{-s}_n\\
&\zeta_{\al}(s|L)\equiv\sum_{n}\lambda^{-s}_n= \(\frac{R_0}{12}\)^{-s}\sum_{n}\(\hat{\lambda}_n-\al\)^{-s},\\
\end{split}\2.
\ee
The eigenvalue of the zero-modes contribution to the calculation should be eliminated, and the $\zeta_{\al}(s|\hat{L})$ function can be expanded as
\be
\label{A2}
\zeta_{\al}(s|L)=\(\frac{R_0}{12}\)^{-s}\[F_{\al}(s)+\sum^{\infty}_{k=0}\frac{\al^k\Gamma(s+k)\hat{G}(s+k)}{k!
\Gamma(s)} \],
\ee
with
\be
\1\{\begin{split}
\label{A3}
&F_{\al}(s)=\sum_{\hat{\lambda}_n<|\al|}\( \hat{\lambda}_n -\al\)^{-s},\hat{F}(s)=\sum_{\hat{\lambda}_n\leqslant|\al|} \hat{\lambda}_n ^{-s}, \\
&\hat{G}(s)=\sum_{\hat{\lambda}_n>|\al|}\hat{\lambda}_n ^{-s}=\hat{\zeta}(s|L)- \hat{F}(s),    F_\al(0)-  \hat{F}(0)=N_0,    \\
\end{split}\2.
\ee
where $N_0$ represents the number of zero-modes. In physics, the zeta-function and its derivative's values at zero play a pivotal role. We consider the Laurent expansion at $s=0$ for this purpose.
\be
\1\{\begin{split}
\label{A4}
&\Gamma(s+k)\hat{\zeta}(s+k)=\frac{\hat{b}_k}{s}+\hat{a_k}+O(s), \Gamma(s+k)\hat{G}(s+k)=\frac{\hat{b}_k}{s}+\hat{a_k}+O(s),\\
&b_0=\hat{b}_{k}-\hat{F(0)}, a_0=\hat{a}_0+\gamma\hat{F}(0), \\
&b_k=\hat{b}_k,a_k=\hat{a}_k-\Gamma(k)\hat{F}(k) ,1\leq k\leq2\\
&b_k=\hat{b}_k=0, \hat{G}(k)=\hat{\zeta}(k|L)- \hat{F}(k) ,k\geq2
\end{split}\2.
\ee
Ultimately, \eqref{A2} can be written as
\be
\label{A5}
\zeta_{\al}(s|L)=\(\frac{R_0}{12}\)^{-s}\[F_{\al}(s)+\sum^{D/2}_{k=0}\(\frac{b_k\al^k}{k!}+s\frac{\(a_k+\gamma b_k\)\al^k}{k!} \)+s\sum_{k>D/2}\frac{\al^k\hat{G}(k)}{k}+O(s^2) \],
\ee
and finally
\be
\1\{\begin{split}
\label{A6}
&\zeta_{\al}(0|L)=\[F_{\al}(0)+\sum^{2}_{k=0}\(\frac{b_k\al^k}{k!}\) \]\\
&\zeta'_{\al}(0|L)=-\zeta_{\al}(0|L)\ln{\frac{R_0}{12}}+\[F'_{\al}(0)+\sum^{D/2}_{k=0}\(\frac{\(a_k+\gamma b_k\)\al^k}{k!} \)+\sum_{k>D/2}\frac{\al^k\hat{G}(k)}{k} \].
\end{split}\2.
\ee
The symbol for the Euler-Mascheroni constant is represented as $\gamma$. Imaginary numbers will appear in the computation results when the model exhibits instability. The eigenvalue $\lambda_n$ and their corresponding degeneracy $g_n$ for the scalar-vector and tensor-type fluctuation operators ($L_i=-\Box_{(i)}-\frac{R_0}{12}q$) acting on de Sitter backgrounds are
\be
\1\{\begin{split}
\label{A7}
&\lambda_n=\frac{R_0}{12}\( \hat{\lambda}_n-\al\),   \hat{\lambda}_n=(n+\nu)^2, g_n=c_1(n+\nu)+c_3(n+\nu)^3\\
&scalar: \nu=\frac{3}{2},\al=\frac{9}{4}+q, c_1=-\frac{1}{12},c_3=\frac{1}{3}\\
&vector:\nu=\frac{5}{2},\al=\frac{13}{4}+q, c_1=-\frac{9}{4},c_3=1\\
&tensor:\nu=\frac{7}{2},\al=\frac{17}{4}+q, c_1=-\frac{125}{12},c_3=\frac{5}{3}.
\end{split}\2.
\ee
We see that $\hat{\zeta}(s)$ is related to well known Hurwitz functions $\zeta_H(s,\nu)$ by
\be
\label{A8}
\begin{split}
\hat{\zeta}(s|L)&=\sum^{\infty}_{n=0}g_n\hat{\lambda}^{-s}_n= \sum^{\infty}_{n=0} \[c_1(n+\nu)^{-(2s-1)}+c_3(n+\nu)^{-(2s-3)} \],\\
&=c_1\zeta_{H}(2s-1,\nu)+c_3\zeta_{H}(2s-3,\nu),
\end{split}
\ee
and
\be
\label{A9}
\begin{split}
\hat{G}(s)&=c_1\zeta_{H}(2s-1,\nu)+c_3\zeta_{H}(2s-3,\nu)-\hat{F}(s)\\
&=c_1\zeta_{H}(2s-1,\nu+m)+c_3\zeta_{H}(2s-3,\nu+m),
\end{split}
\ee
where $m$ is the number of $\hat{\lambda}_n\leq|\al|$, after a series of complex calculation we can obtain
\be
\1\{\begin{split}
\label{A10}
&\hat{b}_0=c_1\zeta_{H}(-1,\nu)+c_3\zeta_{H}(-3,\nu), \hat{b}_1=\frac{c_1}{2},\hat{b}_2=\frac{c_3}{2}\\
&\hat{a}_0=c_1\[2\zeta'_{H}(-1,\nu)-\gamma\zeta_{H}(-1,\nu) \]+c_3\[ 2\zeta'_{H}(-3,\nu)-\gamma\zeta_{H}(-3,\nu)  \]\\
&\hat{a}_1=-c_1\[ \psi(\nu) +\frac{\gamma}{2}\]+c_3\zeta_{H}(-1,\nu)\\
&\hat{a}_2=c_1\zeta_{H}(3,\nu) -c_3\[ \psi(\nu) +\frac{\gamma-1}{2}  \].
\end{split}\2.
\ee
$\psi(s)$ is the logarithmic derivative of Euler's gamma function. we can derive one-loop's calculation by using \eqref{A4},\eqref{A6}. There are three modes for the nonlocal $f(R)$ quantum gravity component of our calculations, which we shall examine each one at later.
\subsection{The tensor case}
The eigenvalue of $L_2$ is
 \be
\label{A11}
\lambda_n=\frac{R_0}{12}\[\(n+\frac{7}{2}\)^2 -\frac{9}{4}\], n=0,1,2...
\ee
Here it can be shown that there is no contribution of zero and negative modes, so $m=0$, for $k\geq3$, we have $F_{\al_2}(s)=0$, $F'_{\al_2}(s)=0$, $\hat{F}(s)=0$, and
\be
\1\{\begin{split}
\label{A12}
&\hat{G}(k)=-\frac{125}{12}\zeta_{H}\(2k-1,\frac{7}{2} \)+\frac{5}{3}\zeta_{H}\(2k-3,\frac{7}{2}\)  \\
&\zeta_{\al_2}(0|L_2)=\sum^{2}_{k=0}\frac{b_k\al^k_2}{k!}=\frac{89}{18}\\
&\zeta'_{\al_2}(0|L_2)=-\frac{89}{18}\ln\frac{R_0}{12}-\frac{125}{6}\zeta'_{H}\(-1,\frac{7}{2} \) +\frac{10}{3}\zeta'_{H}\(-3,\frac{7}{2}\)\\
&+\al_2\[\frac{125}{12}\psi\(\frac{7}{2}\) +\frac{5}{3}\zeta_{H}\(-1,\frac{7}{2} \)    \] +\frac{\al^2_2}{2}\[ -\frac{125}{12}\zeta_{H}\(3,\frac{7}{2} \) -\frac{5}{3}\psi\(\frac{7}{2}\)+\frac{5}{6}  \]\\
&+\sum^{\infty}_{k=3}\frac{\hat{G}(k)\al^k_2}{k}\approx 0.79031-\frac{89}{18}\ln\frac{R_0}{12}      .
\end{split}\2.
\ee
Similarly, the eigenvalue of $\bar{L}_2$ (The off-shell fluctuation operator given by \eqref{36}) is
\be
\1\{\begin{split}
\label{A13}
&\lambda_n=\frac{R_0}{12}\[\(n+\frac{7}{2}\)^2 -\bar{\al}_2\], \bar{\al}_2=\frac{33}{4}+Y,  n=0,1,2...\\
&Y\equiv\frac{-12\(f(R_0)-\phi_0F(R_0)\)}{R_0\( f^{(0)}_{R}-\phi_0F^{(0)}_{R}\)},
\end{split}\2.
\ee
there exist contribution of zero mode and negative mode, depending on the parameter $Y$, and
\be
\1\{\begin{split}
\label{A14}
&\hat{G}(k)=-\frac{125}{12}\zeta_{H}\(2k-1,\frac{7}{2} \)+\frac{5}{3}\zeta_{H}\(2k-3,\frac{7}{2}\)-\hat{F}(k)  \\
&\zeta_{\bar{\al}_2}(0|\bar{L}_2)=F_{\bar{\al}_2}(0)+\sum^{2}_{k=0}\frac{b_k\bar{\al}^k_2}{k!}=\[N_0+\frac{5}{12}\bar{\al} _2^2-\frac{125}{24}\bar{\al}_2+\frac{5}{3} \zeta_H \left(-3,\frac{7}{2}\right)-\frac{125 }{12}\zeta_H \left(-1,\frac{7}{2}\right)\]\\
&\zeta'_{\bar{\al}_2}(0|\bar{L}_2)=\[ N_0+\frac{8383}{576}+\frac{5}{12}\bar{\al} _2^2-\frac{125}{24}\bar{\al}_2\]\ln\frac{R_0}{12}-\frac{125}{6}\zeta'_{H}\(-1,\frac{7}{2} \) +\frac{10}{3}\zeta'_{H}\(-3,\frac{7}{2}\)+F'_{\bar{\al}_2}(0)\\
&+  \bar{\al}_2 \[\frac{125}{12}\psi\(\frac{7}{2}\) +\frac{5}{3}\zeta_{H}\(-1,\frac{7}{2} \)-\hat{F}(1)    \] +\frac{\bar{\al}^2_2}{2}\[ -\frac{125}{12}\zeta_{H}\(3,\frac{7}{2} \) -\frac{5}{3}\psi\(\frac{7}{2}\)+\frac{5}{6}-\hat{F}(2)  \]\\
&+\sum^{\infty}_{k=3}\frac{\hat{G}(k)\bar{\al}_2^k}{k}.
\end{split}\2.
\ee

\subsection{The vector case }
The eigenvalues of $L_1$ is
 \be
\label{A15}
\lambda_n=\frac{R_0}{12}\[\(n+\frac{5}{2}\)^2 -\frac{25}{4}\], n=0,1,2...
\ee
There is a zero-mode $\hat{\lambda}_0=\frac{25}{4}$, so $m=1$, for $k\geq3$, we have $F_{\al_1}(s)=0$, $F'_{\al_1}(s)=0$, $\hat{F}(s)=10\(\frac{25}{4}\)^{-s}$,$N_0=10$, and
\be
\1\{\begin{split}
\label{A16}
&\hat{G}(k)=-\frac{9}{4}\zeta_{H}\(2k-1,\frac{7}{2} \)+\zeta_{H}\(2k-3,\frac{7}{2}\)  \\
&\zeta_{\al_1}(0|L_1)=F_{\al_1}(0)+\sum^{2}_{k=0}\frac{b_k\al^k}{k!}=-\frac{191}{30}\\
&\zeta'_{\al_1}(0|L_1)=\frac{191}{30}\ln\frac{R_0}{12}-\frac{18}{4}\zeta'_{H}\(-1,\frac{5}{2} \) +2\zeta'_{H}\(-3,\frac{5}{2}\)\\
&+\frac{25}{4}\[\frac{9}{4}\psi\(\frac{5}{2}\) +\zeta_{H}\(-1,\frac{5}{2} \)-\frac{8}{5}    \] +\frac{625}{32}\[ -\frac{9}{4}\zeta_{H}\(3,\frac{5}{2} \) -\psi\(\frac{5}{2}\)+\frac{61}{250}  \]\\
 &+\sum^{\infty}_{k=3}\frac{\hat{G}(k)\al^k_1}{k}\approx -18.91+\frac{191}{30}\ln\frac{R_0}{12}      .
\end{split}\2.
\ee
\subsection{The scalar case }
The eigenvalue of $L_{\pm}$ is
 \be
\label{A17}
\lambda_n=\frac{R_0}{12}\[\(n+\frac{3}{2}\)^2 -\al_{\pm}\], n=0,1,2...
\ee
where $\al_{\pm}=\frac{9}{4}+ q_{\pm}$, $q_{\pm}=-\frac{12}{R_0}s_{\pm}$. The value of $\al_{\pm}$ depends on the specific choice of the model. The result may have zero-mode and negative eigenvalue. Finally, we have
 \be
\1\{\begin{split}
\label{A18}
&\hat{G}(k)=-\frac{1}{12}\zeta_{H}\(2k-1,\frac{3}{2} \)+\frac{1}{3}\zeta_{H}\(2k-3,\frac{3}{2}\)-\hat{F}(k)  \\
&\zeta_{\al}(0|L_\pm)=F_{\al}(0)+\sum^{2}_{k=0}\frac{b_k\al_{\pm}^k}{k!}=\[N_0-\frac{1}{12}\zeta_H\(-1,\frac{3}{2}\)+\frac{1}{3}\zeta_{H}\(-3,\frac{3}{2}\)-\frac{1}{24}\al_{\pm}+\frac{1}{12}\al_{\pm}^2          \]\\
&\zeta'_{\al}(0|L_\pm)=-\(N_0-\frac{17}{2880}-\frac{1}{24}\al_{\pm}+\frac{1}{12}\al_{\pm}^2   \)\ln\frac{R_0}{12}\\
&+\frac{1}{3}\[3F'_\al(0)+2\zeta'_{H}\(-3,\frac{3}{2} \) -\frac{1}{2}\zeta'_{H}\(-1,\frac{3}{2}\)       \] -\[72\hat{F}(1)+11-6\psi\(\frac{3}{2}\)\] \frac{\al_{\pm}}{72}\\
&-\[ 12\hat{F}(2)+\zeta_{H}\(3,\frac{3}{2} \)+4\psi\(\frac{3}{2}\) -2 \] \frac{\al_{\pm}^2}{24}\\
&+\sum^{\infty}_{k=3}\frac{\hat{G}(k)\al_{\pm}^k}{k}    .
\end{split}\2.
\ee
The eigenvalue of $L_{0}$ (The off-shell fluctuation operator given by \eqref{36}) is
 \be
\label{A19}
\lambda_n=\frac{R_0}{12}\[\(n+\frac{3}{2}\)^2 -\frac{33}{4}\] , n=0,1,2...
\ee
Then $n=0$ and $n=1$ are smaller than $\al_0$($m$=2), $F_{\al_0}(s)=(-6)^{-s}+5(-2)^{-s}$,  $\hat{F}(s)=\(\frac{9}{4}\)^{-s}+5\(\frac{25}{4}\)^{-s}$($N_0=0$), and
 \be
\1\{\begin{split}
\label{A20}
&\hat{G}(k)=-\frac{1}{12}\zeta_{H}\(2k-1,\frac{7}{2} \)+\frac{1}{3}\zeta_{H}\(2k-3,\frac{7}{2}\)  \\
&\zeta_{\al}(0|L_0)=F_{\al}(0)+\sum^{2}_{k=0}\frac{b_k\al_{\pm}^k}{k!}=\frac{479}{90}\\
&\zeta'_{\al}(0|L_0)=-\frac{479}{90}\ln\frac{R_0}{12}+\frac{1}{3}\[3F'_\al(0)+2\zeta'_{H}\(-3,\frac{3}{2} \) -\frac{1}{2}\zeta'_{H}\(-1,\frac{3}{2}\)       \]\\
&-\[72\hat{F}(1)+11-6\psi\(\frac{3}{2}\)\] \frac{\al_{0}}{72}-\[ 12\hat{F}(2)+\zeta_{H}\(3,\frac{3}{2} \)+4\psi\(\frac{3}{2}\) -2 \] \frac{\al_{0}^2}{24}\\
&+\sum^{\infty}_{k=3}\frac{\hat{G}(k)\al_{\pm}^k}{k}    .
\end{split}\2.
\ee

\appendix
\bibliographystyle{unsrt}
\bibliography{oneloop.v4}

\begin{thebibliography}{10}

\bibitem{SupernovaCosmologyProject:1998vns}
S.~Perlmutter et~al.
\newblock {Measurements of $\Omega$ and $\Lambda$ from 42 high redshift
  supernovae}.
\newblock {\em Astrophys. J.}, 517:565--586, 1999.

\bibitem{SupernovaSearchTeam:1998fmf}
Adam~G. Riess et~al.
\newblock {Observational evidence from supernovae for an accelerating universe
  and a cosmological constant}.
\newblock {\em Astron. J.}, 116:1009--1038, 1998.

\bibitem{SNLS:2005qlf}
P.~Astier et~al.
\newblock {The Supernova Legacy Survey: Measurement of $\Omega_M$,
  $\Omega_\Lambda$ and ${\cal w}$ from the first year data set}.
\newblock {\em Astron. Astrophys.}, 447:31--48, 2006.

\bibitem{SDSS:2003eyi}
Max Tegmark et~al.
\newblock {Cosmological parameters from SDSS and WMAP}.
\newblock {\em Phys. Rev. D}, 69:103501, 2004.

\bibitem{WMAP:2010qai}
E.~Komatsu et~al.
\newblock {Seven-Year Wilkinson Microwave Anisotropy Probe (WMAP) Observations:
  Cosmological Interpretation}.
\newblock {\em Astrophys. J. Suppl.}, 192:18, 2011.

\bibitem{SDSS:2004kqt}
Uros Seljak et~al.
\newblock {Cosmological parameter analysis including SDSS Ly-alpha forest and
  galaxy bias: Constraints on the primordial spectrum of fluctuations, neutrino
  mass, and dark energy}.
\newblock {\em Phys. Rev. D}, 71:103515, 2005.

\bibitem{SDSS:2005xqv}
Daniel~J. Eisenstein et~al.
\newblock {Detection of the Baryon Acoustic Peak in the Large-Scale Correlation
  Function of SDSS Luminous Red Galaxies}.
\newblock {\em Astrophys. J.}, 633:560--574, 2005.

\bibitem{Jain:2003tba}
Bhuvnesh Jain and Andy Taylor.
\newblock {Cross-correlation tomography: measuring dark energy evolution with
  weak lensing}.
\newblock {\em Phys. Rev. Lett.}, 91:141302, 2003.

\bibitem{Kilbinger:2008gk}
M.~Kilbinger et~al.
\newblock {Dark energy constraints and correlations with systematics from
  CFHTLS weak lensing, SNLS supernovae Ia and WMAP5}.
\newblock {\em Astron. Astrophys.}, 497:677, 2009.

\bibitem{t1974one}
Gerard t~Hooft and MJG1974AnIHP Veltman.
\newblock One-loop divergencies in the theory of gravitation.
\newblock In {\em Annales de l'IHP Physique th{\'e}orique}, volume~20, pages
  69--94, 1974.

\bibitem{Deser:1974cz}
Stanley Deser and P.~van Nieuwenhuizen.
\newblock {One Loop Divergences of Quantized Einstein-Maxwell Fields}.
\newblock {\em Phys. Rev. D}, 10:401, 1974.

\bibitem{Deser:1974xq}
Stanley Deser, Hung-Sheng Tsao, and P.~van Nieuwenhuizen.
\newblock {One Loop Divergences of the Einstein Yang-Mills System}.
\newblock {\em Phys. Rev. D}, 10:3337, 1974.

\bibitem{Buchbinder:1992rb}
I.~L. Buchbinder, S.~D. Odintsov, and I.~L. Shapiro.
\newblock {\em {Effective action in quantum gravity}}.
\newblock 1992.

\bibitem{Codello:2007bd}
Alessandro Codello, Roberto Percacci, and Christoph Rahmede.
\newblock {Ultraviolet properties of f(R)-gravity}.
\newblock {\em Int. J. Mod. Phys. A}, 23:143--150, 2008.

\bibitem{Machado:2007ea}
Pedro~F. Machado and Frank Saueressig.
\newblock {On the renormalization group flow of f(R)-gravity}.
\newblock {\em Phys. Rev. D}, 77:124045, 2008.

\bibitem{Codello:2008vh}
Alessandro Codello, Roberto Percacci, and Christoph Rahmede.
\newblock {Investigating the Ultraviolet Properties of Gravity with a Wilsonian
  Renormalization Group Equation}.
\newblock {\em Annals Phys.}, 324:414--469, 2009.

\bibitem{Knorr:2019atm}
Benjamin Knorr, Chris Ripken, and Frank Saueressig.
\newblock {Form Factors in Asymptotic Safety: conceptual ideas and
  computational toolbox}.
\newblock {\em Class. Quant. Grav.}, 36(23):234001, 2019.

\bibitem{Cognola:2005de}
Guido Cognola, Emilio Elizalde, Shin'ichi Nojiri, Sergei~D. Odintsov, and
  Sergio Zerbini.
\newblock {One-loop f(R) gravity in de Sitter universe}.
\newblock {\em JCAP}, 02:010, 2005.

\bibitem{Ruf:2017bqx}
Michael~S. Ruf and Christian~F. Steinwachs.
\newblock {One-loop divergences for $f(R)$ gravity}.
\newblock {\em Phys. Rev. D}, 97(4):044049, 2018.

\bibitem{Alvarez-Gaume:2015rwa}
Luis Alvarez-Gaume, Alex Kehagias, Costas Kounnas, Dieter L\"ust, and Antonio
  Riotto.
\newblock {Aspects of Quadratic Gravity}.
\newblock {\em Fortsch. Phys.}, 64(2-3):176--189, 2016.

\bibitem{Salam:1978fd}
Abdus Salam and J.~A. Strathdee.
\newblock {Remarks on High-energy Stability and Renormalizability of Gravity
  Theory}.
\newblock {\em Phys. Rev. D}, 18:4480, 1978.

\bibitem{Masuda:1976qg}
Naohiko Masuda and Richard~M. Weiner.
\newblock {Energy Distribution of Secondaries in Proton-Nucleus Collisions at
  Very High-Energies}.
\newblock {\em Phys. Lett. B}, 70:77--82, 1977.

\bibitem{Tomboulis:1983sw}
E.~T. Tomboulis.
\newblock {Unitarity in Higher Derivative Quantum Gravity}.
\newblock {\em Phys. Rev. Lett.}, 52:1173, 1984.

\bibitem{PhysRevD.16.953}
K.~S. Stelle.
\newblock Renormalization of higher-derivative quantum gravity.
\newblock {\em Phys. Rev. D}, 16:953--969, Aug 1977.

\bibitem{Tomboulis:1977jk}
E.~Tomboulis.
\newblock {1/N Expansion and Renormalization in Quantum Gravity}.
\newblock {\em Phys. Lett. B}, 70:361--364, 1977.

\bibitem{Tomboulis:1980bs}
E.~Tomboulis.
\newblock {Renormalizability and Asymptotic Freedom in Quantum Gravity}.
\newblock {\em Phys. Lett. B}, 97:77--80, 1980.

\bibitem{Antoniadis:1986tu}
Ignatios Antoniadis and E.~T. Tomboulis.
\newblock {Gauge Invariance and Unitarity in Higher Derivative Quantum
  Gravity}.
\newblock {\em Phys. Rev. D}, 33:2756, 1986.

\bibitem{Modesto:2017sdr}
Leonardo Modesto and Les\l{}aw Rachwa\l{}.
\newblock {Nonlocal quantum gravity: A review}.
\newblock {\em Int. J. Mod. Phys. D}, 26(11):1730020, 2017.

\bibitem{Belgacem:2017cqo}
Enis Belgacem, Yves Dirian, Stefano Foffa, and Michele Maggiore.
\newblock {Nonlocal gravity. Conceptual aspects and cosmological predictions}.
\newblock {\em JCAP}, 03:002, 2018.

\bibitem{Koshelev:2016xqb}
Alexey~S. Koshelev, Leonardo Modesto, Leslaw Rachwal, and Alexei~A.
  Starobinsky.
\newblock {Occurrence of exact $R^2$ inflation in non-local UV-complete
  gravity}.
\newblock {\em JHEP}, 11:067, 2016.

\bibitem{Dimitrijevic:2022fhj}
Ivan Dimitrijevic, Branko Dragovich, Zoran Rakic, and Jelena Stankovic.
\newblock {Nonlocal de Sitter gravity and its exact cosmological solutions}.
\newblock {\em JHEP}, 12:054, 2022.

\bibitem{Deser:2007jk}
Stanley Deser and R.~P. Woodard.
\newblock {Nonlocal Cosmology}.
\newblock {\em Phys. Rev. Lett.}, 99:111301, 2007.

\bibitem{Koivisto:2008dh}
Tomi~S. Koivisto.
\newblock {Newtonian limit of nonlocal cosmology}.
\newblock {\em Phys. Rev. D}, 78:123505, 2008.

\bibitem{NOJIRI2008821}
Shin'ichi Nojiri and Sergei~D. Odintsov.
\newblock Modified non-local-f(r) gravity as the key for the inflation and dark
  energy.
\newblock {\em Physics Letters B}, 659(4):821--826, 2008.

\bibitem{Bamba:2012ky}
Kazuharu Bamba, Shinichi Nojiri, Sergei~D. Odintsov, and Misao Sasaki.
\newblock {Screening of cosmological constant for De Sitter Universe in
  non-local gravity, phantom-divide crossing and finite-time future
  singularities}.
\newblock {\em Gen. Rel. Grav.}, 44:1321--1356, 2012.

\bibitem{Zhang:2011uv}
Ying-li Zhang and Misao Sasaki.
\newblock {Screening of cosmological constant in non-local cosmology}.
\newblock {\em Int. J. Mod. Phys. D}, 21:1250006, 2012.

\bibitem{Elizalde:2011su}
E.~Elizalde, E.~O. Pozdeeva, and S.~Yu. Vernov.
\newblock {De Sitter Universe in Non-local Gravity}.
\newblock {\em Phys. Rev. D}, 85:044002, 2012.

\bibitem{Nojiri:2010pw}
Shin'ichi Nojiri, Sergei~D. Odintsov, Misao Sasaki, and Ying-li Zhang.
\newblock {Screening of cosmological constant in non-local gravity}.
\newblock {\em Phys. Lett. B}, 696:278--282, 2011.

\bibitem{Zhang:2016ykx}
Ying-li Zhang, Kazuya Koyama, Misao Sasaki, and Gong-Bo Zhao.
\newblock {Acausality in Nonlocal Gravity Theory}.
\newblock {\em JHEP}, 03:039, 2016.

\bibitem{Nojiri:2007uq}
Shin'ichi Nojiri and Sergei~D. Odintsov.
\newblock {Modified non-local-F(R) gravity as the key for the inflation and
  dark energy}.
\newblock {\em Phys. Lett. B}, 659:821--826, 2008.

\bibitem{Jhingan:2008ym}
S.~Jhingan, S.~Nojiri, S.~D. Odintsov, M.~Sami, I~Thongkool, and S.~Zerbini.
\newblock {Phantom and non-phantom dark energy: The Cosmological relevance of
  non-locally corrected gravity}.
\newblock {\em Phys. Lett. B}, 663:424--428, 2008.

\bibitem{Nojiri:2019dio}
Shin'ichi Nojiri, S.~D. Odintsov, and V.~K. Oikonomou.
\newblock {Ghost-free non-local $F(R)$ Gravity Cosmology}.
\newblock {\em Phys. Dark Univ.}, 28:100541, 2020.

\bibitem{Cognola:2007vq}
Guido Cognola, Monica Gastaldi, and Sergio Zerbini.
\newblock {On the stability of a class of modified gravitational models}.
\newblock {\em Int. J. Theor. Phys.}, 47:898--910, 2008.

\bibitem{Pozdeeva:2019agu}
Ekaterina~O. Pozdeeva, Mohammad Sami, Alexey~V. Toporensky, and Sergey~Yu.
  Vernov.
\newblock {Stability analysis of de Sitter solutions in models with the
  Gauss-Bonnet term}.
\newblock {\em Phys. Rev. D}, 100(8):083527, 2019.

\bibitem{Esposito1997}
Giampiero Esposito, Alexander~Yu. Kamenshchik, and Giuseppe Pollifrone.
\newblock {\em Effective Action in Quantum Field Theory}, pages 51--79.
\newblock Springer Netherlands, Dordrecht, 1997.

\bibitem{Bytsenko:1994bc}
Andrei~A. Bytsenko, Guido Cognola, Luciano Vanzo, and Sergio Zerbini.
\newblock {Quantum fields and extended objects in space-times with constant
  curvature spatial section}.
\newblock {\em Phys. Rept.}, 266:1--126, 1996.

\bibitem{Vassilevich:2003xt}
D.~V. Vassilevich.
\newblock {Heat kernel expansion: User's manual}.
\newblock {\em Phys. Rept.}, 388:279--360, 2003.

\bibitem{Dowker:1975tf}
J.~S. Dowker and Raymond Critchley.
\newblock {Effective Lagrangian and Energy Momentum Tensor in de Sitter Space}.
\newblock {\em Phys. Rev. D}, 13:3224, 1976.

\bibitem{Fradkin:1983mq}
E.~S. Fradkin and Arkady~A. Tseytlin.
\newblock {One Loop Effective Potential in Gauged O(4) Supergravity}.
\newblock {\em Nucl. Phys. B}, 234:472, 1984.

\bibitem{Shahidi:2018smw}
Shahab Shahidi, Farid Charmchi, Zahra Haghani, and Leila Shahkarami.
\newblock {Modified gravity one-loop partition function}.
\newblock {\em Eur. Phys. J. C}, 78(10):833, 2018.

\end{thebibliography}

\end{document}